\documentclass[twocolumn,english,aps,prd,nofootinbib]{revtex4}
\usepackage[T1]{fontenc}
\usepackage[utf8]{inputenc}
\usepackage{amsmath}
\usepackage{graphicx}
\usepackage{amssymb}
\usepackage{esint}

\usepackage{mciteplus}

\newcommand{\q}[1]{q\nobreakdash-\hspace{0pt}}
\newcommand{\cdott}{\!\cdot\!} 
\newcommand{\as}{\alpha_{\mathrm s}} 
\newcommand{\CF}{c_{\mathrm F}} 
\newcommand{\CA}{c_{\mathrm A}} 
\newcommand{\TR}{T_{\mathrm R}} 
\newcommand{\NF}{N_{\mathrm f}} 
\newcommand{\mur}{\mu_{\mathrm r}} 

\begin{document}

\title{General Mass Scheme for Jet Production in DIS}
\author{P. Kotko}
\email{piotr.kotko@ifj.edu.pl}
\author{W. Slominski}
\email{wojtek.slominski@uj.edu.pl}
\affiliation{M. Smoluchowski Institute of Physics, Jagiellonian University, Reymonta 4, 
30-059 Krak\'ow, Poland}

\begin{abstract}
We propose a method for calculating DIS jet production cross sections in QCD at
NLO accuracy with consistent treatment of heavy quarks. The scheme
relies on the dipole subtraction method for jets, which we extend to all
possible initial state splittings with heavy partons,
so that the Aivazis-Collins-Olness-Tung
massive collinear factorization scheme (ACOT) can be applied. As a
first check of the formalism we recover the ACOT result for the heavy
quark structure function using a dedicated Monte Carlo program.
\end{abstract}
\maketitle

\section{Introduction}
The contributions of heavy quarks to the inclusive DIS processes can be consistently
described using so called general mass schemes (GM) (see e.g. \citep{Thorne:2008xf}
for a review). The GM schemes are designed to solve the problem occurring in
perturbative QCD when two or more large scales change significantly
with respect to each other. In DIS the large scales are the
virtuality of exchanged boson, $Q^{2}$ and the mass $m_{\mathbf{Q}}$
of the heavy quark $\mathbf{Q}$ (assume for simplicity that there is
only one heavy quark). Let us thus consider the ratio $\rho=m_{\mathbf{Q}}^{2}/Q^{2}$.
There are two distinct regions, $\rho\simeq 0$ and $\rho\sim 1$,
where different approaches are used. 
In the first region $\mathbf{Q}$ is treated as massless and the factorization theorem is used to resum the resulting collinear singularities into the parton distribution function (PDF) for $\mathbf{Q}$.
In the other
region ($\rho\sim 1$) the effects of the heavy quark mass are fully taken into account, order by order in the perturbation expansion.
In particular, the arising powers of
$\as \log\rho$
are finite and no resummation is required.
This approach is referred to as `fixed order'.

The GM schemes provide a description over the whole range of $\rho$,
by means of different resummation methods of the potentially large $(\as \log\rho)^n$ terms.
There are several solutions of this kind 
\citep{Aivazis:1993pi,Thorne:1997ga,*Thorne:1997uu,Kramer:2000hn,Forte:2010ta}
which are used in the description of inclusive data. For the purpose
of our investigation the ACOT scheme \citep{Aivazis:1993pi} is of
particular interest as it actually is a factorization theorem with
heavy quarks, proved to all orders in QCD \citep{Collins:1998rz}.

The GM schemes have been so far formulated and used for \emph{inclusive}
DIS processes only. The purpose of this letter is to report on a new
ACOT-based GM scheme for jet production in DIS. All necessary calculations
were made in \citep{Kotko_phdthesis,*Kotko:2012kw} and a detailed analysis will
be presented in a forthcoming paper.

Analysis of the jet production processes is a key tool to study various
aspects of perturbative QCD. On the theoretical side the calculations
are much more involved than in the inclusive case. This is due to
infrared (IR) singularities that appear at intermediate stages of
calculations but eventually get canceled in the physical cross sections.
The problems originate in the fact that the phase space is not complete
as the final state partons have to produce distinct jets. 
Thus, in
practice, only Monte Carlo (MC) methods are applicable here and special
methods for canceling IR singularities between real and virtual corrections
have to be used. 

One of the exact solutions is the dipole subtraction method (DSM) \citep{Catani:1996vz,Frixione:1997np}.
It was constructed initially for massless quarks only, but later a
complete extension for massive quarks in the \emph{final state}
was made \citep{Phaf:2001gc,Catani:2002hc}. The method made use of
Ref. \citep{Dittmaier:1999mb}, where photon radiation off (possibly
massive) fermions was considered, also in the initial state. 

It is important to note that the GM schemes deal with initial state splitting
processes, where two of the participating partons are heavy quarks.
None of the methods mentioned above take this into account completely.
In Ref. \citep{Catani:2002hc} there are no massive partons in the initial
state splittings, while Ref. \citep{Dittmaier:1999mb} considers only
$\mathbf{Q}\rightarrow\mathbf{Q}\gamma$ emissions.

The method presented in this paper consists in two basic elements:
i) dipole subtraction method extended to all possible QCD splitting
processes with heavy quarks, ii) the ACOT factorization scheme
for initial state 
\emph{quasi-collinear singularities}.
The last notion
corresponds to a situation, where both transverse momentum and
mass of the emitted parton tend to zero in a uniform way \citep{Catani:2000ef}.
In what follows we shall refer to these quasi-collinear singularities as
`\q-singularities' for brevity.

In Section \ref{sec:DSM} we briefly recall DSM and describe our extensions.
Next, in Section \ref{sec:Factorization} we address the problem of
IR sensitive logarithms in the framework of the ACOT scheme. Finally,
in Section \ref{sec:Example} we present a test of our method by an explicit
numerical MC calculation of the charm structure function at NLO.

\section{Dipole subtraction method with heavy quarks}
\label{sec:DSM}


Let us consider a calculation of $n$-jet cross section to NLO accuracy (our notation
is close to that of Ref. \citep{Catani:2002hc}).
First, the LO contribution is (schematically) 
\begin{equation}
\sigma_{n}^{\left(\mathrm{LO}\right)}=\mathcal{N}\,\sum_{a}f_{a}\otimes\,\int d\Phi_{n,a}\,\left|\mathcal{M}_{n,a}\right|^{2}F_{n,a},
\end{equation}
where $f_{a}$ is a PDF for parton $a$, the symbol $\otimes$ denotes
convolution and $\mathcal{N}$ is an appropriate normalization factor. The $n$-particle
phase space (PS) is denoted by $d\Phi_{n,a}$, while $\mathcal{M}_{n,a}$
is a tree-level matrix element (ME) with $n$ partons in the final
state. The jet function $F_{n,a}$ produces the pertinent observable
upon integration over the phase space. It has the property that $F_{n+1,a}=F_{n,a}$
in the singular regions of phase space (IR safety condition). Consider
now the NLO contribution. Within DSM it reads
\begin{widetext}
\begin{equation}
\sigma_{n}^{\left(\mathrm{NLO}\right)} \! = \mathcal{N}\,\sum_{a}
f_{a}\otimes\Bigg\{
\int \!\! d\Phi_{n+1,a}\,\left[\left|\mathcal{M}_{n+1,a}\right|^{2}F_{n+1,a}-\mathcal{D}_{n,a}F_{n,a}\right]
+\int \!\! d\Phi_{n,a}\,\left[\mathcal{M}_{n,a}^{\left(\mathrm{loop}\right)}
+\int\!\! d\phi_{a}\,\mathcal{D}_{n,a}-\mathcal{C}_{n,a}\right] F_{n,a}\Bigg\}.
\label{eq:dipole_general}
\end{equation}
\end{widetext} 

The notation $\mathcal{M}_{n,a}^{\left(\mathrm{loop}\right)}$ corresponds
to virtual corrections to $\left|\mathcal{M}_{n,a}\right|^{2}$. The
auxiliary function $\mathcal{D}_{n,a}$ (called 
`dipole function'
 in the following) is chosen in such a way that: i) it exactly mimics soft
and \q-singularities of $\left|\mathcal{M}_{n+1,a}\right|^{2}$ without
double counting them in the overlap region \citep{Catani:2002hc},
ii) it can be analytically integrated over the one-particle subspace
$d\phi_{a}$ defined 
symbolically
 by the relation $d\Phi_{n+1,a}=d\Phi_{n,a}\otimes d\phi_{a}$.
The novelty here is that also the initial state \q-singularities
are considered when constructing the dipole functions. 
By construction,
the first square bracket (real emissions) is integrable in four 
dimensions, thanks to the properties of jet functions. In order to leave the
cross section unchanged we have to add the same dipole contribution
as we have subtracted. This time, however, we use the integrated dipole
function using its property ii). 
In general, this integrated dipole contains IR poles (in dimensional
regularization), which eventually
get canceled against the poles in $\mathcal{M}_{n}^{\left(\mathrm{loop}\right)}$.
The resulting expression is, however, still not IR safe and the remaining \q-singularities
have to be factorized out. It is achieved by subtracting
following `counterterms'
\begin{equation}
\mathcal{C}_{n,a}=\sum_{b}\mathcal{F}_{ab}\otimes\left|\mathcal{M}_{n,b}\right|^{2}.\label{eq:coll_sub_term}\end{equation}
The quantities $\mathcal{F}_{ab}$ are the renormalized densities of partons
$b$ inside parton $a$. We shall come back to these objects later
in Section \ref{sec:Factorization}.

Let us now define the dipole function more precisely. It is given by
the sum of three (for DIS) classes of contributions
\begin{multline}
\mathcal{D}_{n,a}\left(p_{a};\left\{ p_{l}\right\} _{l=1}^{n+1}\right)=\sum_{i,\underset{\!\!\!\!{\scriptstyle j\neq i}}{j=1}}^{n+1}\bigg\{ D_{i,j,a}^{\left(A\right)}\left(\tilde{p}_{\underline{ai}};\mathfrak{p}_{A}\right)\\
+D_{i,j,a}^{\left(B\right)}\left(\tilde{p}_{a};\mathfrak{p}_{B}\right)+\sum_{\underset{{\scriptstyle k\neq i,j}}{k=1}}^{n+1}D_{i,j,k}^{\left(C\right)}\left(p_{a};\mathfrak{p}_{C}\right)\bigg\},
\label{eq:Dipconst_D}
\end{multline}
where $p_{a}$ is the initial state parton momentum and $p_{l}$ with
$l=1,\ldots n+1$ are final state momenta satisfying momentum conservation
$q+p_{a}=\sum_{l}p_{l}$. The three classes of dipoles are:
$\left(A\right)$
initial state emitter with final state spectator (IE-FS),
$\left(B\right)$
final state emitter with initial state spectator (FE-IS),
and $\left(C\right)$
final state emitter with final state spectator (FE-FS).
The `emitter'
and `spectator' partons are defined here following Ref. \citep{Catani:1996vz}.
The FE-FS case $\left(C\right)$ does not involve initial states and
was completely covered in \citep{Catani:2002hc}, thus we do not consider
it further.

\begin{figure}
\centerline{%
\includegraphics[width=0.95\columnwidth]{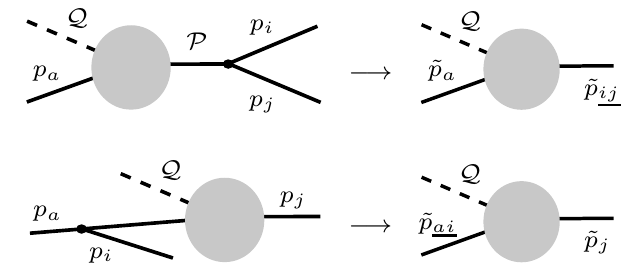}%
}
\caption{Kinematics for FE-IS (upper) and IE-FS (lower).\label{fig:Fig1}}
\end{figure}

The remaining dipoles in (\ref{eq:Dipconst_D}) read
\begin{subequations}
\label{eq:Dipconst_D2}
\begin{multline}
D_{i,j,a}^{\left(A\right)}\left(\tilde{p}_{\underline{ai}};\mathfrak{p}_{A}\right)=-\frac{1}{\left(p_{a}-p_{i}\right)^{2}-m_{\underline{ai}}^{2}}\,\frac{1}{x}\\
\big<\mathcal{M}_{n}\left(\tilde{p}_{\underline{ai}};\mathfrak{p}_{A}\right)\big|\frac{\hat{T}_{j}\cdott\hat{T}_{\underline{ai}}}{\hat{T}_{\underline{ai}}^{2}}\,\hat{V}_{a\rightarrow\underline{ai}\, i,\, j}^{\left(A\right)}\big|\mathcal{M}_{n}\left(\tilde{p}_{\underline{ai}};\mathfrak{p}_{A}\right)\big>,
\label{eq:Dipconst_D_IEFS}
\end{multline}
\begin{multline}
D_{i,j,a}^{\left(B\right)}\left(\tilde{p}_{a};\mathfrak{p}_{B}\right)=-\frac{1}{\left(p_{i}+p_{j}\right)^{2}-m_{\underline{ij}}^{2}}\,\frac{1}{x}\\
\big<\mathcal{M}_{n}\left(\tilde{p}_{a};\mathfrak{p}_{B}\right)\big|\frac{\hat{T}_{a}\cdott\hat{T}_{\underline{ij}}}{\hat{T}_{\underline{ij}}^{2}}\,\hat{V}_{\underline{ij}\rightarrow i\, j,\, a}^{\left(B\right)}\big|\mathcal{M}_{n}\left(\tilde{p}_{a};\mathfrak{p}_{B}\right)\big>,
\label{eq:Dipconst_D_FEIS}
\end{multline}
\end{subequations}
where $\big|\mathcal{M}_{n}\big>$ is a vector in helicity and color
space such that $\left|\mathcal{M}_{n}\right|^{2}=\big<\mathcal{M}_{n}\big|\mathcal{M}_{n}\big>$
is a matrix element squared, summed/averaged over colors and spins
(see \citep{Catani:1996vz} for details). The kinematics is depicted
in Fig.~\ref{fig:Fig1}. For IE-FS the partons $a,\, i$ are replaced
by a single parton $\underline{ai}$, while for FE-IS $i,\, j$ are
replaced by $\underline{ij}$. Also the spectators acquire new momenta.
All the new momenta in $\mathcal{M}_{n}$ are constructed to be on
shell and satisfy the momentum conservation, also away from the IR
limits. We adorn these new momenta with tilde. The resulting sets
of final state momenta in $\mathcal{M}_{n}$ are denoted by $\mathfrak{p}_{A}$ 
and $\mathfrak{p}_{B}$, respectively.

The $x$ variable is the Sudakov longitudinal fraction of $\tilde{p}_{a}$
or $\tilde{p}_{\underline{ai}}$ along $p_{a}$ and the color operators
$\hat{T}_{l}$ pick up relevant color factors and account for color
correlations. The dipole splitting functions $\hat{V}$ act in the helicity
space and account for spin correlations. In the quasi-collinear limit
they tend to massive generalizations of splitting matrices, while
in the soft limit they collapse to eikonal factors.

\subsection{Dipole splitting functions}

Here we present the realization of $\hat{V}$ in a general massive
case. It is worth mentioning that our results extend those of Ref.
\citep{Catani:2002hc}, and coincide 
in suitable limits with the partially massless cases
considered there.

First, let us define the tilded momenta as follows
\begin{subequations}
\begin{gather}
\textrm{IE-FS:}\quad\tilde{p}_{j}^{\mu}=\tilde{w}_{A}\,\mathcal{P}^{\mu}-\tilde{u}_{A}\, p_{a}^{\mu},\quad\tilde{p}_{\underline{ai}}^{\mu}=\tilde{p}_{j}^{\mu}-\mathcal{Q}^{\mu},
\label{eq:Dipkin_FEIS_3}\\
\textrm{FE-IS:}\quad\tilde{p}_{\underline{ij}}^{\mu}=\tilde{w}_{B}\,\mathcal{P}^{\mu}-\tilde{u}_{B}\, p_{a}^{\mu},\quad\tilde{p}_{a}^{\mu}=\tilde{p}_{\underline{ij}}^{\mu}-\mathcal{Q}^{\mu},
\label{eq:Dipkin_FEIS_4}
\end{gather}
\end{subequations}
where $\mathcal{P}=p_{i}+p_{j}$, $\mathcal{Q}=\mathcal{P}-p_{a}$
(Fig.~\ref{fig:Fig1}) and $\tilde{u}$, $\tilde{w}$ variables are
calculated from the on-shell conditions. In the soft limit $\tilde{u}\rightarrow0$,
$\tilde{w}\rightarrow1$. Moreover $\tilde{u}\rightarrow1-x$ in the
quasi-collinear limit. 

Let us now list our dipole splitting functions starting with IE-FS
case. For a gluon emission $\mathbf{Q}\rightarrow\mathbf{Q}g$ (or
$\overline{\mathbf{Q}}\rightarrow\overline{\mathbf{Q}}g$) we have
(with $m_{\mathbf{Q}}\equiv m_{a}=m_{\underline{ai}}$, $m_{i}=0$)
\begin{multline}
\hat{V}_{\mathbf{Q}\rightarrow\mathbf{Q}g,\, j}^{\left(A\right)}=8\pi\mur^{2\varepsilon}\alpha_{s}\CF\\
\left[\frac{2}{\tilde{u}_{A}\tilde{v}_{j}^{2}+\tilde{z}}+\left(1-\varepsilon\right)\tilde{u}_{A}-2-\frac{1-\tilde{u}_{A}}{\tilde{z}}\,\frac{m_{\mathbf{Q}}^{2}}{\mathcal{P}_{a}}\right],\end{multline}
where $\tilde{z}=p_{i}\cdot p_{a}/\mathcal{P}_{a}$ with $\mathcal{P}_{a}=\mathcal{P}\cdot p_{a}$.
Further, $\mur$ is a mass scale needed in $D=4-2\varepsilon$ dimensions,
and $\tilde{v}_{j}=\sqrt{1-\left(m_{a}m_{j}/\tilde{\gamma}\right)^{2}}$
with $\tilde{\gamma}=\tilde{p}_{j}\cdot p_{a}$. Above (and in what
follows) the unit matrix in helicity space is suppressed. 

For the $g\rightarrow\mathbf{Q}\overline{\mathbf{Q}}$ splitting we
get \begin{multline}
\hat{V}_{g\rightarrow\mathbf{Q}\overline{\mathbf{Q}},\, j}^{\left(A\right)}=8\pi\mur^{2\varepsilon}\alpha_{s}\TR\\
\left[1-\frac{1}{1-\varepsilon}\,\left(2\tilde{u}_{A}\left(1-\tilde{u}_{A}\right)-\frac{1-\tilde{u}_{A}}{\tilde{z}}\,\frac{m_{\mathbf{Q}}^{2}}{\mathcal{P}_{a}}\right)\right],\label{eq:Dipsplit_IEFS_g_QQ_V}\end{multline}
with the configuration of masses $m_{a}=0$, $m_{\mathbf{Q}}\equiv m_{\underline{ai}}=m_{i}$.
Note that although the initial state parton is massless, this splitting
is not present in \citep{Dittmaier:1999mb,Catani:2002hc}.

Finally for $\mathbf{Q}\rightarrow g\mathbf{Q}$, where a
heavy quark is
radiated, we have (with $m_{a}=m_{i}=m_{\mathbf{Q}}$, $m_{\underline{ai}}=0$)\begin{multline}
\left(\hat{V}_{\mathbf{Q}\rightarrow g\mathbf{Q},\, j}^{\left(A\right)}\right)^{\mu\nu}=8\pi\mur^{2\varepsilon}\alpha_{s}\CF\left(1-\varepsilon\right)\\
\left[-g^{\mu\nu}\left(1-\tilde{u}_{A}\right)+\frac{4}{1-\tilde{u}_{A}}\,\frac{\mathcal{V}_{A}^{\mu}\mathcal{V}_{A}^{\nu}}{2\tilde{z}\mathcal{P}_{a}-m_{\mathbf{Q}}^{2}}\right].\label{eq:Dipsplit_IEFS_Q_gQ_V}\end{multline}
The correlation tensor $\mathcal{V}_{A}^{\mu}\mathcal{V}_{A}^{\nu}$
is defined by\begin{multline}
\mathcal{V}_{A}^{\mu}=\left(1-\tilde{z}\right)p_{i}^{\mu}-\tilde{z}p_{j}^{\mu}\\
+\left[m_{\mathbf{Q}}^{2}-m_{j}^{2}+\mathcal{P}^{2}\left(1-2\tilde{z}\right)\right]\frac{1-\tilde{w}_{A}}{2\tilde{p}_{\underline{ai}}\cdot\mathcal{P}}\,\mathcal{P}^{\mu}.\end{multline}
It fulfills transversality relation $\tilde{p}_{\underline{ai}}\cdot\mathcal{V}_{A}=0$.

Now let us turn to the FE-IS case. Here the difference with \citep{Catani:2002hc}
is that the initial state spectator can be massive. For $\mathbf{Q}\rightarrow\mathbf{Q}g$
with $m_{i}=0$, $m_{j}=m_{\underline{ij}}=m_{\mathbf{Q}}$ we have
\begin{multline}
\hat{V}_{\mathbf{Q}\rightarrow\mathbf{Q}g,\, a}^{\left(B\right)}=8\pi\mur^{2\varepsilon}\alpha_{s}\CF\\
\left[\frac{2}{\tilde{u}_{B}\tilde{v}_{\mathbf{Q}}^{2}+\tilde{z}}+\left(1-\varepsilon\right)\tilde{z}-2-\frac{m_{\mathbf{Q}}^{2}}{p_{i}\cdot p_{j}}\right],\label{eq:Dipsplit_FEIS_Q_Qg_V}\end{multline}

Further for $g\rightarrow\mathbf{Q}\overline{\mathbf{Q}}$ with $m_{\underline{ij}}=0$,
$m_{i}=m_{j}=m_{\mathbf{Q}}$ we define \begin{equation}
\left(\hat{V}_{g\rightarrow\mathbf{Q}\overline{\mathbf{Q}},\, a}^{\left(B\right)}\right)^{\mu\nu}=8\pi\mur^{2\varepsilon}\alpha_{s}\TR\left(-g^{\mu\nu}-4\frac{\mathcal{V}_{B}^{\mu}\mathcal{V}_{B}^{\nu}}{\mathcal{P}^{2}}\right),\label{eq:Dipspli_FEIS_g_QQ_Vgen}\end{equation}
with\begin{equation}
\mathcal{V}_{B}^{\mu}=\tilde{z}p_{i}^{\mu}-\left(1-\tilde{z}\right)p_{j}^{\mu}-\frac{\tilde{u}_{B}m_{a}^{2}}{2\tilde{w}_{B}\mathcal{P}_{a}}\left(p_{i}^{\mu}-p_{j}^{\mu}\right)\end{equation}
satisfying $\tilde{p}_{\underline{ij}}\cdot\mathcal{V}_{B}=0$. 

Finally for the $g\rightarrow gg$ splitting we have \begin{multline}
\left(\hat{V}_{g\rightarrow gg,\, a}^{\left(B\right)}\right)^{\mu\nu}=16\pi\mur^{2\varepsilon}\alpha_{s}\CA\Bigg[-g^{\mu\nu}\Bigg(\frac{1}{1-\tilde{z}+\tilde{u}_{B}}\\
+\frac{1}{\tilde{z}+\tilde{u}_{B}}-2\Bigg)+2\left(1-\varepsilon\right)\frac{\mathcal{V}_{B}^{\mu}\mathcal{V}_{B}^{\nu}}{\mathcal{P}^{2}}\Bigg].\label{eq:Dipsplit_FEIS_g_gg_V}\end{multline}
Despite the massless splitting the above function gains some massive
factors from spectator via the $\mathcal{V}_{B}^{\mu}$ vectors.

\subsection{Dipole integration}

The phase space factorization required in Eq. (\ref{eq:dipole_general}) is
actually realized as a convolution in $\tilde{u}$ treated as a free
variable. In this context we denote it by $u$ (without tilde). The
construction requires dropping on-shell conditions for $\tilde{p}_{j}$
in IE-FS or $\tilde{p}_{\underline{ij}}$ in FE-IS cases and fixing
two invariants $\mathcal{X}$, $\mathcal{Y}$, such that we can determine
$\tilde{w}=\tilde{w}\left(u,\mathcal{X},\mathcal{Y}\right)$. Let
us denote the corresponding off-shell vectors as $\tilde{p}_{j}\left(u,\mathcal{X},\mathcal{Y}\right)\equiv\tilde{p}_{j}\left(u\right)$,
$\tilde{p}_{\underline{ij}}\left(u,\mathcal{X},\mathcal{Y}\right)\equiv\tilde{p}_{\underline{ij}}\left(u\right)$.

Our version of PS factorization formula has the form (for IE-FS case)\begin{multline}
d\Phi_{n+1}\left(q,p_{a};\left\{ p_{l}\right\} _{l=1}^{n+1}\right)=\\
\int\!\! du\, d\Phi_{n}\left(\mathcal{Q},\tilde{p}_{\underline{ai}}\left(u\right);\mathfrak{p}_{A}\right)\, d\phi_{a\rightarrow\underline{ai}\, i,\, j}^{\left(A\right)}\left(u\right),\label{eq:PS_fact}\end{multline}
where the arguments of the reduced phase space $d\Phi_{n}$ 
are the incoming
and outgoing
momenta (separated by semicolon). 
The subspace measure
is \begin{multline}
d\phi_{a\rightarrow\underline{ai}\, i,\, j}^{\left(A\right)}\left(u\right)=v^{3-D}\,\left(\mathcal{P}^{2}\right)^{\frac{D}{2}-2}\mathcal{J}\,\\
\left[\left(\tilde{z}_{+}-\tilde{z}\right)\left(\tilde{z}-\tilde{z}_{-}\right)\right]^{\frac{D}{2}-2}\,\frac{d\Omega_{D-2}d\tilde{z}}{4\left(2\pi\right)^{D-1}},\label{eq:PSfactoriz_FEIS_dphi}\end{multline}
where $v=\sqrt{1-m_{a}^{2}\mathcal{P}^{2}/\mathcal{P}_{a}^{2}}$ and
$d\Omega_{D-2}$ is a solid angle on transverse hyperplane in the $\mathcal{Q}$-$p_{a}$
CM system. The jacobian reads

\begin{equation}
\mathcal{J}=\left|\frac{\partial\tilde{p}_{j}^{2}\left(u\right)}{\partial u}\right|_{\mathcal{X},\mathcal{Y}},
\end{equation}
with subscripts corresponding to variables kept fixed during differentiation.
The bounds on $\tilde{z}$ are $\tilde{z}_{\pm}=\frac{1}{2}\left(1+\overline{m}_{i}^{2}-\overline{m}_{j}^{2}\pm2v\overline{p}\right)$,
where 
$\overline{p}=\frac{1}{2}\sqrt{1-2\left(\overline{m}_{i}^{2}+\overline{m}_{j}^{2}\right)+\left(\overline{m}_{i}^{2}-\overline{m}_{j}^{2}\right)^{2}}$
and the barred masses are rescaled by $\mathcal{P}^{2}$.

An analogous formula can be obtained for the FE-IS case. One has to replace
the set $\mathfrak{p}_A$ by $\mathfrak{p}_B$ and make replacements:
$\tilde{p}_{\underline{ai}}\leftrightarrow\tilde{p}_{a}$ in $d\Phi_{n}$
and $\tilde{p}_{j}\rightarrow\tilde{p}_{\underline{ij}}$ in the jacobian
$\mathcal{J}$.

The choice of invariants $\mathcal{X}$, $\mathcal{Y}$ affects the
PS generation in $d\Phi_{n}$ and `plus-distributions' handling
in the integrals over $d\phi$. 
In the following we take $\mathcal{X}=\tilde{\gamma}$,
$\mathcal{Y}=\mathcal{P}_{a}$.

We note that in the desired integral $\int d\phi\, D$, with $D$ given
in (\ref{eq:Dipconst_D2}),
only $\hat{V}$ and the propagators 
can depend on $\tilde{z}$. Further,
it can be shown that, thanks to the gauge invariance and properties of
$\mathcal{V}_{A,B}^{\mu}$, the helicity correlations in (\ref{eq:Dipconst_D2})
vanish after integration. Therefore the integral over $d\Omega_{D-2}$ is trivial and we
are left with a color correlated matrix element and a scalar function,
which e.g. for IE-FS reads
\begin{equation}
I_{a\rightarrow\underline{ai}\, i,\, j}^{\left(A\right)}\left(u\right)=-\int\!\! d\phi_{a\rightarrow\underline{ai}\, i,\, j}^{\left(A\right)}\,\,\frac{\left\langle \hat{V}_{a\rightarrow\underline{ai}\, i,\, j}^{\left(A\right)}\right\rangle }{\left(p_{a}-p_{i}\right)^{2}-m_{\underline{ai}}^{2}}\,,
\end{equation}
where $\left\langle \ldots\right\rangle $ denotes average over helicities
in $D$ dimensions. The formula for FE-IS is analogous.

All the integrals have been calculated analytically in $D$ dimensions
\citep{Kotko_phdthesis,*Kotko:2012kw},
and the details will be given in a forthcoming paper. 

As an example,
let us present the result for the IE-FS $g\rightarrow\mathbf{Q}\overline{\mathbf{Q}}$ dipole
integral
as we shall make use of it in Section \ref{sec:Factorization}.
In the $D=4$ limit we get
\begin{multline}
{I}_{g\rightarrow\mathbf{Q}\overline{\mathbf{Q}},\, j}^{(A)}\left(u\right)
=\frac{\eta_{\mathcal{J}}^{2}}{2\eta_{\mathcal{P}_{a}}^{4}}\,
\\ \times 
\Bigg\{\eta_{\mathcal{P}_{a}}^{2}P_{gq}\left(u\right)\,\log\frac{\eta_{\mathbf{Q}}^{2}-\eta_{j}^{2}+\eta_{\mathcal{P}^{2}}^{2}\left(1+2\overline{p}\right)}{\eta_{\mathbf{Q}}^{2}-\eta_{j}^{2}+\eta_{\mathcal{P}^{2}}^{2}\left(1-2\overline{p}\right)}\\
+\frac{8\,\TR\,\eta_{\mathbf{Q}}^{2}\, \eta_{\mathcal{P}^{2}}^{4}\left(1-u\right)\overline{p}}{\left(\eta_{\mathbf{Q}}^{2}-\eta_{j}^{2}+\eta_{\mathcal{P}^{2}}^{2}\right)^{2}-4\eta_{\mathcal{P}^{2}}^{4}\overline{p}^{2}}\Bigg\},
\label{eq:Intdip_IEFS_g_QQ_4D}
\end{multline}
where $P_{gq}\left(u\right)=\TR\left[1-2u\left(1-u\right)\right]$
is the standard splitting function,
$\eta_j^{2}=m_{j}^{2}/2\tilde{\gamma}$,
$\eta_{\mathbf{Q}}^{2}=m_{\mathbf{Q}}^{2}/2\tilde{\gamma}$
and other $\eta^{2}_X = X/2\tilde{\gamma}$ for
$X = \mathcal{P}_{a}, \mathcal{P}^2, \mathcal{J}$.
Note that in the above formula a non-trivial dependence on $u$ is hidden in 
$\overline{p}$ and $\eta^{2}_X$.

\section{Factorization of (quasi-)singularities}

\label{sec:Factorization}

In general, the integrated dipole functions contain IR poles, which
get canceled upon including virtual corrections. 
There are no further singularities in the massive case. Nevertheless, some IR sensitive
terms (\q-singularities) still remain in the IE-FS dipoles,
and we factorize them out by subtracting counterterms
(\ref{eq:coll_sub_term})
according to the ACOT scheme, as we explain below.

Let us consider
the integrated IE-FS dipole
in the limit of vanishing mass
of the emitter or emitted parton (the spectator $j$ can remain massive).
For instance, using Eq. (\ref{eq:Intdip_IEFS_g_QQ_4D}),
we obtain
\begin{multline}
I_{g\rightarrow\mathbf{Q}\overline{\mathbf{Q}},\, j}^{\left(A\right)}\left(u\right)
=
\frac{\alpha_{s}}{2\pi}\Bigg[P_{gq}\left(u\right)\Bigg(\log\frac{u^{2}}{u+\eta_{j}^{2}}-\log\eta_{\mathbf{Q}}^{2}\Bigg)\\
+2\TR\, u\left(1-u\right)\Bigg]+\mathcal{O}\left(\eta_{\mathbf{Q}}^{2}\right),
\end{multline}

The counterterms $\mathcal{C}_{n,a}$
are given in terms of the partonic densities
$\mathcal{F}_{ab}$, 
which are defined via matrix elements of bilocal operators on the light-cone
and can be calculated perturbatively. 
The emerging UV singularities have to be renormalized which, in turn, specifies the evolution kernels for PDFs
(see \citep{Collins:2011zzd} for a review).

According to the ACOT scheme, this standard procedure can also be used with massive quarks \citep{Collins:1998rz}. In particular, 
one can still define
partonic PDFs
and renormalize them using $\overline{\mathrm{MS}}$,
assuring standard DGLAP evolution equations (actually one uses the composite
CWZ renormalization scheme \citep{Collins:1978wz}).
For asymptotically large $Q^2$, the above construction leads to the standard $\overline{\mathrm{MS}}$ calculation. On the other hand, when
$Q^2\simeq m^2_{\mathbf{Q}}$ the $\mathit{C}_{n,a}$ counterterms 
approximately cancel 
the $\sigma^{(\mathrm{LO})}$
contribution, 
basically resulting in the `fixed order' description~\citep{Aivazis:1993pi}.

Using the Feynman rules for PDFs \citep{Collins:1981uw,Collins:2011zzd}
one can show that in the $\overline{\mathrm{MS}}$ scheme
\begin{equation}
\mathcal{F}_{g\mathbf{Q}}\left(z\right)=\frac{\alpha_{s}}{2\pi}\,
 \TR\,
\left(1-2z\left(1-z\right)\right)\,
\log \frac{\mur^{2}}{m_{\mathbf{Q}}^{2}}
\, ,
\end{equation}
\begin{multline}
\mathcal{\mathcal{F}}_{\mathbf{Q}g}\left(z\right)=\frac{\alpha_{s}}{2\pi}\,
 \CF\,\frac{1+\left(1-z\right)^{2}}{z}
\\
\left[\log \frac{\mur^{2}}{m_{\mathbf{Q}}^{2}} -2\log z-1\right]
\end{multline}
\begin{multline}
\mathcal{\mathcal{F}}_{\mathbf{Q}\mathbf{Q}}\left(z\right)=\frac{\alpha_{s}}{2\pi}\,
 \CF\\
\left\{ \frac{1+z^{2}}{1-z}
\left[\log \frac{\mur^{2}}{m_{\mathbf{Q}}^{2}} -2\log\left(1-z\right)-1\right]
\right\} _{+},
\end{multline}
\begin{multline}
\mathcal{F}_{gg}\left(z\right) = \frac{\alpha_{s}}{2\pi}\, 
\\
\Bigg\{2\CA\,\left[\left(\frac{1}{1-z}\right)_{+}+\frac{1-z}{z}-1+z\left(1-z\right)\right]\\
+\delta\left(1-z\right)\left(\frac{11}{6}\CA-\frac{2}{3}\NF\, \TR-\frac{2}{3}\TR\log\frac{\mur^{2}}{m_{\mathbf{Q}}^{2}}\right)\Bigg\}.
\end{multline}
where 
$z$ is the fractional momentum of
parton $b$. 

We have checked that using 
$\mathcal{C}_{n,a}$ with 
the above $\mathcal{F}_{ab}$
(intermediate steps require
usage of color conservation for $\hat{T}_{j}$ operators) we get IR
safe dipoles. Moreover, in the considered limit we get the same dipoles
as those of Ref. \citep{Catani:2002hc} with the standard massless collinear
subtraction term (i.e. with $\mathcal{F}_{ab}\left(z\right)=-\alpha_{s}P_{ab}\left(z\right)/2\pi\varepsilon$).

There is one comment in order. 
Certain ambiguities \citep{Collins:1998rz} of the ACOT scheme
stimulated construction of the
so called 
Simplified ACOT (S-ACOT)
scheme \citep{Kramer:2000hn}, where all initial state masses are
set to zero. This scheme is convenient for higher order calculations
\citep{Stavreva:2012bs}, which otherwise are very cumbersome. 
Our
approach is however different as it operates on more exclusive quantities,
hence we resolve the ambiguities differently. Namely, we set the initial
state masses to zero in $\sigma^{(\mathrm{LO})}$
and in the
$\mathcal{C}_{n,a}$ contribution to $\sigma^{(\mathrm{NLO})}$.
This results in massless integration limits in the corresponding convolutions,
in analogy to the standard DGLAP evolution equations.
This approach is not strictly required by our formalism,
but provides a better transition to the `fixed order' description,
cf. solid and dotted lines in Fig.~\ref{fig:GeneralMass_Example_2}.
We point out that we \emph{do
not set} any other masses to zero, neither for initial nor for
the final states. If, however, one wants to use the S-ACOT scheme for jets, then only
the dipoles for the $g\rightarrow\mathbf{Q}\overline{\mathbf{Q}}$ initial
state splitting are needed.

\section{Numerical tests}

\label{sec:Example}

We have partially implemented our method in a dedicated C++ program.
The MC integration and event generation is accomplished using the
FOAM engine \citep{Jadach:2002kn}.

In order to test our approach, we have used our program to numerically calculate
a quantity which can be obtained independently, namely
the charm structure function $F_{2c}$.

The LO contribution comes from
$\gamma\mathbf{Q}\rightarrow\mathbf{Q}$
(plus the same with $\overline{\mathbf{Q}}$).
The real NLO corrections are $\gamma\mathbf{Q}\rightarrow\mathbf{Q}g$ and $\gamma g\rightarrow\mathbf{Q}\overline{\mathbf{Q}}$,
while
the virtual ones
are given e.g. in \citep{Kretzer:1998ju}.

In order to perform the numerical integration
we need three distinct dipoles 
$D_{\mathbf{Q},\overline{\mathbf{Q}},g}^{\left(A\right)}=D_{\overline{\mathbf{Q}},\mathbf{Q},g}^{\left(A\right)}$,
$D_{g,\mathbf{Q},\mathbf{Q}}^{\left(A\right)}=D_{g,\overline{\mathbf{Q}},\overline{\mathbf{Q}}}^{\left(A\right)}$,
$D_{g,\mathbf{Q},\mathbf{Q}}^{\left(B\right)}=D_{g,\overline{\mathbf{Q}},\overline{\mathbf{Q}}}^{\left(B\right)}$
as well as their integrals. Since there are two IE-FS dipoles we need two collinear
subtraction terms with corresponding $\mathcal{F}_{g\mathbf{Q}}$
and $\mathcal{F}_{\mathbf{Q\mathbf{Q}}}$.

\begin{figure}
\centerline{%
\includegraphics[width=0.99\columnwidth]{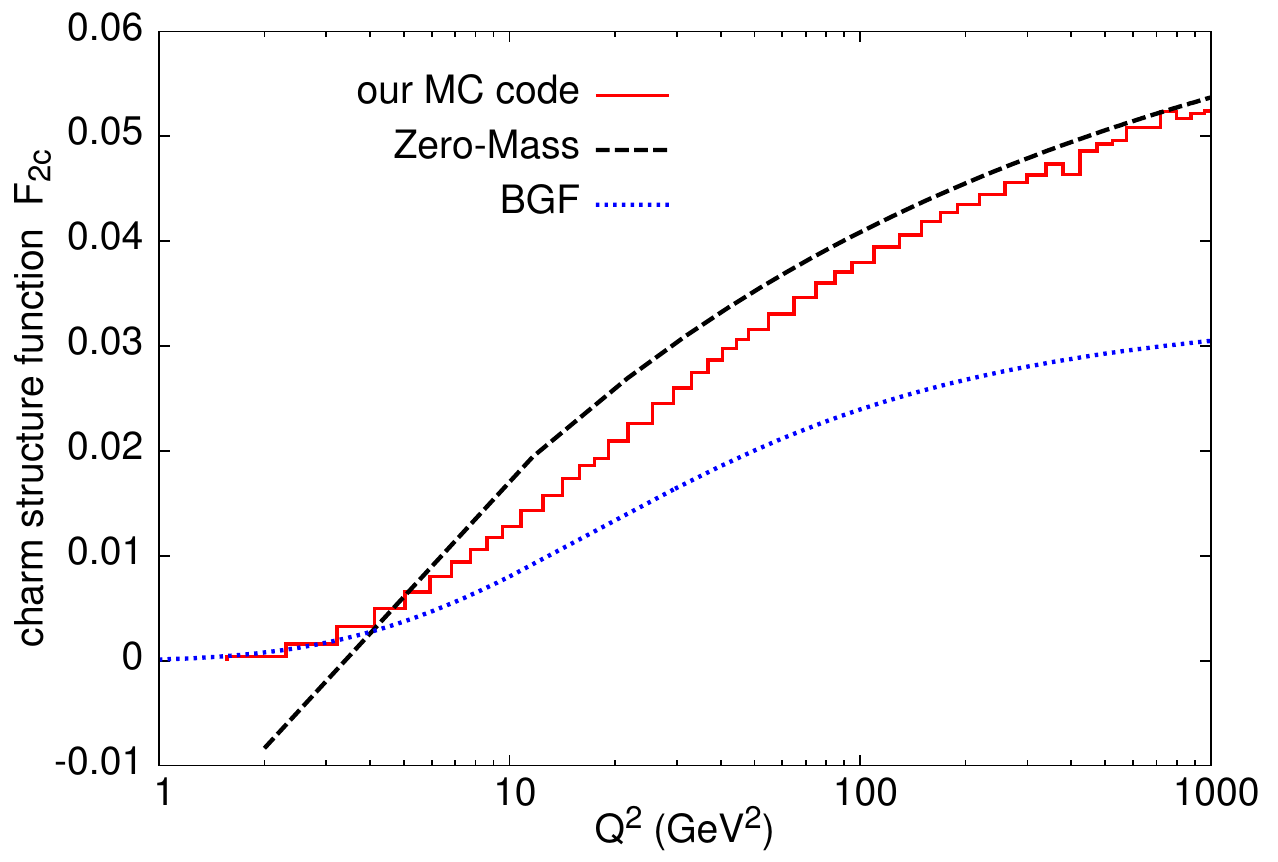}%
}
\caption{The charm structure function, $F_{2c}$, calculated in different schemes: 
GM from our MC code at NLO (red solid line), Zero-Mass at NLO (black dashed line),
Fixed Order (BGF) at ${\cal O}(\as)$ (blue dotted line). The calculations were
done using CTEQ5 PDFs at $x_{\rm B}=0.05$. 
\label{fig:GeneralMass_Example_2}}
\end{figure}

We observe, that the soft poles coming from virtual corrections cancel
against the sum of soft poles coming from IE-FS and FE-IS
dipoles. 
The result of numerical integration perfectly agrees with the semi-analytical 
formula calculated independently according to \citep{Aivazis:1993pi,Kretzer:1998ju}.
In Fig.~\ref{fig:GeneralMass_Example_2} we show
that our result indeed interpolates between the massless NLO calculation
at high $Q^{2}$ and the fixed order, ${\cal O}(\as)$, calculation at low $Q^{2}$. Let us
stress that this result is obtained by the MC integration of a fully differential
cross section (2).

This is clearly the first simplest test.
It verifies
only a part of the dipoles and collinear counterterms.
More tests are due in course of the code development.

\section{Summary}

In the present paper we have briefly described --- without going into
technical details --- a general mass scheme for jet production in
DIS. The scheme is based on the ACOT factorization theorem which is
proved to all orders. The proof is constructed for structure functions,
but it actually holds for all IR safe quantities. 
This entitles us to 
use this factorization scheme for jets, provided we have a suitable
dipole subtraction method that can deal with heavy quarks in the initial
state. We have constructed such a method, generalizing the existing results.
We have checked that 
this method removes all potential
collinear singularities as desired, 
and that the remaining dipoles coincide with those
of the massless calculation in $\overline{\mathrm{MS}}$.

All the crucial calculations are already performed \citep{Kotko_phdthesis,*Kotko:2012kw}
and the details shall be presented in a separate publication.
The computer code suitable for the jet production in DIS is under construction.

An extension to hadron-hadron collisions is, in principle, straightforward.
Formally, one only needs to introduce an
additional class of dipoles for the initial state emitter and initial
state spectator.
Nonetheless, we are aware of challenges arising at higher orders (above NLO)
--- like the feasibility of dipole subtraction method or non-cancellation
of soft singularities when two initial state partons are massive.
These issues have been widely discussed in literature, 
but practical procedures still have to be analyzed.

\begin{acknowledgments}
The work is supported by the Polish National Science 
Center grant no.\ DEC-2011/01/B/ST2/03643.
\end{acknowledgments}

\bibliographystyle{apsrevM}
\bibliography{HeavyFlavours}

\ifx\mcitethebibliography\mciteundefinedmacro
\PackageError{apsrevM.bst}{mciteplus.sty has not been loaded}
{This bibstyle requires the use of the mciteplus package.}\fi
\begin{mcitethebibliography}{21}
\expandafter\ifx\csname natexlab\endcsname\relax\def\natexlab#1{#1}\fi
\expandafter\ifx\csname bibnamefont\endcsname\relax
  \def\bibnamefont#1{#1}\fi
\expandafter\ifx\csname bibfnamefont\endcsname\relax
  \def\bibfnamefont#1{#1}\fi
\expandafter\ifx\csname citenamefont\endcsname\relax
  \def\citenamefont#1{#1}\fi
\expandafter\ifx\csname url\endcsname\relax
  \def\url#1{\texttt{#1}}\fi
\expandafter\ifx\csname urlprefix\endcsname\relax\def\urlprefix{URL }\fi
\providecommand{\bibinfo}[2]{#2}
\providecommand{\eprint}[2][]{\url{#2}}

\bibitem[{\citenamefont{Thorne and Tung}(2008)}]{Thorne:2008xf}
\bibinfo{author}{\bibfnamefont{R.~S.} \bibnamefont{Thorne}} \bibnamefont{and}
  \bibinfo{author}{\bibfnamefont{W.~K.} \bibnamefont{Tung}}
  (\bibinfo{year}{2008}), \eprint{0809.0714}\relax
\mciteBstWouldAddEndPuncttrue
\mciteSetBstMidEndSepPunct{\mcitedefaultmidpunct}
{\mcitedefaultendpunct}{\mcitedefaultseppunct}\relax
\EndOfBibitem
\bibitem[{\citenamefont{Aivazis et~al.}(1994)\citenamefont{Aivazis, Collins,
  Olness, and Tung}}]{Aivazis:1993pi}
\bibinfo{author}{\bibfnamefont{M.~A.~G.} \bibnamefont{Aivazis}},
  \bibinfo{author}{\bibfnamefont{J.~C.} \bibnamefont{Collins}},
  \bibinfo{author}{\bibfnamefont{F.~I.} \bibnamefont{Olness}},
  \bibnamefont{and} \bibinfo{author}{\bibfnamefont{W.-K.} \bibnamefont{Tung}},
  \bibinfo{journal}{Phys. Rev.} \textbf{\bibinfo{volume}{D50}},
  \bibinfo{pages}{3102} (\bibinfo{year}{1994}), \eprint{hep-ph/9312319}\relax
\mciteBstWouldAddEndPuncttrue
\mciteSetBstMidEndSepPunct{\mcitedefaultmidpunct}
{\mcitedefaultendpunct}{\mcitedefaultseppunct}\relax
\EndOfBibitem
\bibitem[{\citenamefont{Thorne and
  Roberts}(1998{\natexlab{a}})}]{Thorne:1997ga}
\bibinfo{author}{\bibfnamefont{R.~S.} \bibnamefont{Thorne}} \bibnamefont{and}
  \bibinfo{author}{\bibfnamefont{R.~G.} \bibnamefont{Roberts}},
  \bibinfo{journal}{Phys. Rev.} \textbf{\bibinfo{volume}{D57}},
  \bibinfo{pages}{6871} (\bibinfo{year}{1998}{\natexlab{a}}),
  \eprint{hep-ph/9709442}\relax
\mciteBstWouldAddEndPuncttrue
\mciteSetBstMidEndSepPunct{\mcitedefaultmidpunct}
{\mcitedefaultendpunct}{\mcitedefaultseppunct}\relax
\EndOfBibitem
\bibitem[{\citenamefont{Thorne and
  Roberts}(1998{\natexlab{b}})}]{Thorne:1997uu}
\bibinfo{author}{\bibfnamefont{R.~S.} \bibnamefont{Thorne}} \bibnamefont{and}
  \bibinfo{author}{\bibfnamefont{R.~G.} \bibnamefont{Roberts}},
  \bibinfo{journal}{Phys. Lett.} \textbf{\bibinfo{volume}{B421}},
  \bibinfo{pages}{303} (\bibinfo{year}{1998}{\natexlab{b}}),
  \eprint{hep-ph/9711223}\relax
\mciteBstWouldAddEndPuncttrue
\mciteSetBstMidEndSepPunct{\mcitedefaultmidpunct}
{\mcitedefaultendpunct}{\mcitedefaultseppunct}\relax
\EndOfBibitem
\bibitem[{\citenamefont{Kramer et~al.}(2000)\citenamefont{Kramer, Olness, and
  Soper}}]{Kramer:2000hn}
\bibinfo{author}{\bibfnamefont{M.}~\bibnamefont{Kramer}},
  \bibinfo{author}{\bibfnamefont{F.~I.} \bibnamefont{Olness}},
  \bibnamefont{and} \bibinfo{author}{\bibfnamefont{D.~E.} \bibnamefont{Soper}},
  \bibinfo{journal}{Phys. Rev.} \textbf{\bibinfo{volume}{D62}},
  \bibinfo{pages}{096007} (\bibinfo{year}{2000}), \eprint{hep-ph/0003035}\relax
\mciteBstWouldAddEndPuncttrue
\mciteSetBstMidEndSepPunct{\mcitedefaultmidpunct}
{\mcitedefaultendpunct}{\mcitedefaultseppunct}\relax
\EndOfBibitem
\bibitem[{\citenamefont{Forte et~al.}(2010)\citenamefont{Forte, Laenen, Nason,
  and Rojo}}]{Forte:2010ta}
\bibinfo{author}{\bibfnamefont{S.}~\bibnamefont{Forte}},
  \bibinfo{author}{\bibfnamefont{E.}~\bibnamefont{Laenen}},
  \bibinfo{author}{\bibfnamefont{P.}~\bibnamefont{Nason}}, \bibnamefont{and}
  \bibinfo{author}{\bibfnamefont{J.}~\bibnamefont{Rojo}},
  \bibinfo{journal}{Nucl.Phys.} \textbf{\bibinfo{volume}{B834}},
  \bibinfo{pages}{116} (\bibinfo{year}{2010}), \eprint{1001.2312}\relax
\mciteBstWouldAddEndPuncttrue
\mciteSetBstMidEndSepPunct{\mcitedefaultmidpunct}
{\mcitedefaultendpunct}{\mcitedefaultseppunct}\relax
\EndOfBibitem
\bibitem[{\citenamefont{Collins}(1998)}]{Collins:1998rz}
\bibinfo{author}{\bibfnamefont{J.~C.} \bibnamefont{Collins}},
  \bibinfo{journal}{Phys. Rev.} \textbf{\bibinfo{volume}{D58}},
  \bibinfo{pages}{094002} (\bibinfo{year}{1998}), \eprint{hep-ph/9806259}\relax
\mciteBstWouldAddEndPuncttrue
\mciteSetBstMidEndSepPunct{\mcitedefaultmidpunct}
{\mcitedefaultendpunct}{\mcitedefaultseppunct}\relax
\EndOfBibitem
\bibitem[{\citenamefont{Kotko}(2012)}]{Kotko_phdthesis}
\bibinfo{author}{\bibfnamefont{P.}~\bibnamefont{Kotko}}, Ph.D. thesis,
  \bibinfo{school}{Jagiellonian Univ.} (\bibinfo{year}{2012})\relax
\mciteBstWouldAddEndPuncttrue
\mciteSetBstMidEndSepPunct{\mcitedefaultmidpunct}
{\mcitedefaultendpunct}{\mcitedefaultseppunct}\relax
\EndOfBibitem
\bibitem[{\citenamefont{Kotko and Slominski}(2012)}]{Kotko:2012kw}
\bibinfo{author}{\bibfnamefont{P.}~\bibnamefont{Kotko}} \bibnamefont{and}
  \bibinfo{author}{\bibfnamefont{W.}~\bibnamefont{Slominski}}
  (\bibinfo{year}{2012}), \bibinfo{note}{prepared for 20th International
  Workshop on Deep Inelastic Scattering and Related Subjects (DIS 2012), Bonn,
  Germany, 26--30 Apr 2012.}, \eprint{1206.3517}\relax
\mciteBstWouldAddEndPuncttrue
\mciteSetBstMidEndSepPunct{\mcitedefaultmidpunct}
{\mcitedefaultendpunct}{\mcitedefaultseppunct}\relax
\EndOfBibitem
\bibitem[{\citenamefont{Catani and Seymour}(1997)}]{Catani:1996vz}
\bibinfo{author}{\bibfnamefont{S.}~\bibnamefont{Catani}} \bibnamefont{and}
  \bibinfo{author}{\bibfnamefont{M.~H.} \bibnamefont{Seymour}},
  \bibinfo{journal}{Nucl. Phys.} \textbf{\bibinfo{volume}{B485}},
  \bibinfo{pages}{291} (\bibinfo{year}{1997}), \bibinfo{note}{erratum: {\em
  ibid.} {\bf B510}:503-504,1998; [arXiv:hep-ph/9605323v3] includes changes
  from the Erratum}, \eprint{hep-ph/9605323}\relax
\mciteBstWouldAddEndPuncttrue
\mciteSetBstMidEndSepPunct{\mcitedefaultmidpunct}
{\mcitedefaultendpunct}{\mcitedefaultseppunct}\relax
\EndOfBibitem
\bibitem[{\citenamefont{Frixione}(1997)}]{Frixione:1997np}
\bibinfo{author}{\bibfnamefont{S.}~\bibnamefont{Frixione}},
  \bibinfo{journal}{Nucl. Phys.} \textbf{\bibinfo{volume}{B507}},
  \bibinfo{pages}{295} (\bibinfo{year}{1997}), \eprint{hep-ph/9706545}\relax
\mciteBstWouldAddEndPuncttrue
\mciteSetBstMidEndSepPunct{\mcitedefaultmidpunct}
{\mcitedefaultendpunct}{\mcitedefaultseppunct}\relax
\EndOfBibitem
\bibitem[{\citenamefont{Phaf and Weinzierl}(2001)}]{Phaf:2001gc}
\bibinfo{author}{\bibfnamefont{L.}~\bibnamefont{Phaf}} \bibnamefont{and}
  \bibinfo{author}{\bibfnamefont{S.}~\bibnamefont{Weinzierl}},
  \bibinfo{journal}{JHEP} \textbf{\bibinfo{volume}{0104}}, \bibinfo{pages}{006}
  (\bibinfo{year}{2001}), \eprint{hep-ph/0102207}\relax
\mciteBstWouldAddEndPuncttrue
\mciteSetBstMidEndSepPunct{\mcitedefaultmidpunct}
{\mcitedefaultendpunct}{\mcitedefaultseppunct}\relax
\EndOfBibitem
\bibitem[{\citenamefont{Catani et~al.}(2002)\citenamefont{Catani, Dittmaier,
  Seymour, and Trocsanyi}}]{Catani:2002hc}
\bibinfo{author}{\bibfnamefont{S.}~\bibnamefont{Catani}},
  \bibinfo{author}{\bibfnamefont{S.}~\bibnamefont{Dittmaier}},
  \bibinfo{author}{\bibfnamefont{M.~H.} \bibnamefont{Seymour}},
  \bibnamefont{and}
  \bibinfo{author}{\bibfnamefont{Z.}~\bibnamefont{Trocsanyi}},
  \bibinfo{journal}{Nucl. Phys.} \textbf{\bibinfo{volume}{B627}},
  \bibinfo{pages}{189} (\bibinfo{year}{2002}), \eprint{hep-ph/0201036}\relax
\mciteBstWouldAddEndPuncttrue
\mciteSetBstMidEndSepPunct{\mcitedefaultmidpunct}
{\mcitedefaultendpunct}{\mcitedefaultseppunct}\relax
\EndOfBibitem
\bibitem[{\citenamefont{Dittmaier}(2000)}]{Dittmaier:1999mb}
\bibinfo{author}{\bibfnamefont{S.}~\bibnamefont{Dittmaier}},
  \bibinfo{journal}{Nucl. Phys.} \textbf{\bibinfo{volume}{B565}},
  \bibinfo{pages}{69} (\bibinfo{year}{2000}), \eprint{hep-ph/9904440}\relax
\mciteBstWouldAddEndPuncttrue
\mciteSetBstMidEndSepPunct{\mcitedefaultmidpunct}
{\mcitedefaultendpunct}{\mcitedefaultseppunct}\relax
\EndOfBibitem
\bibitem[{\citenamefont{Catani et~al.}(2001)\citenamefont{Catani, Dittmaier,
  and Trocsanyi}}]{Catani:2000ef}
\bibinfo{author}{\bibfnamefont{S.}~\bibnamefont{Catani}},
  \bibinfo{author}{\bibfnamefont{S.}~\bibnamefont{Dittmaier}},
  \bibnamefont{and}
  \bibinfo{author}{\bibfnamefont{Z.}~\bibnamefont{Trocsanyi}},
  \bibinfo{journal}{Phys. Lett.} \textbf{\bibinfo{volume}{B500}},
  \bibinfo{pages}{149} (\bibinfo{year}{2001}), \eprint{hep-ph/0011222}\relax
\mciteBstWouldAddEndPuncttrue
\mciteSetBstMidEndSepPunct{\mcitedefaultmidpunct}
{\mcitedefaultendpunct}{\mcitedefaultseppunct}\relax
\EndOfBibitem
\bibitem[{\citenamefont{Collins}(2011)}]{Collins:2011zzd}
\bibinfo{author}{\bibfnamefont{J.}~\bibnamefont{Collins}},
  \emph{\bibinfo{title}{{Foundations of perturbative QCD}}},
  vol.~\bibinfo{volume}{32} (\bibinfo{publisher}{Cambridge Univ. Press},
  \bibinfo{year}{2011})\relax
\mciteBstWouldAddEndPuncttrue
\mciteSetBstMidEndSepPunct{\mcitedefaultmidpunct}
{\mcitedefaultendpunct}{\mcitedefaultseppunct}\relax
\EndOfBibitem
\bibitem[{\citenamefont{Collins et~al.}(1978)\citenamefont{Collins, Wilczek,
  and Zee}}]{Collins:1978wz}
\bibinfo{author}{\bibfnamefont{J.~C.} \bibnamefont{Collins}},
  \bibinfo{author}{\bibfnamefont{F.}~\bibnamefont{Wilczek}}, \bibnamefont{and}
  \bibinfo{author}{\bibfnamefont{A.}~\bibnamefont{Zee}},
  \bibinfo{journal}{Phys. Rev.} \textbf{\bibinfo{volume}{D18}},
  \bibinfo{pages}{242} (\bibinfo{year}{1978})\relax
\mciteBstWouldAddEndPuncttrue
\mciteSetBstMidEndSepPunct{\mcitedefaultmidpunct}
{\mcitedefaultendpunct}{\mcitedefaultseppunct}\relax
\EndOfBibitem
\bibitem[{\citenamefont{Collins and Soper}(1982)}]{Collins:1981uw}
\bibinfo{author}{\bibfnamefont{J.~C.} \bibnamefont{Collins}} \bibnamefont{and}
  \bibinfo{author}{\bibfnamefont{D.~E.} \bibnamefont{Soper}},
  \bibinfo{journal}{Nucl.Phys.} \textbf{\bibinfo{volume}{B194}},
  \bibinfo{pages}{445} (\bibinfo{year}{1982})\relax
\mciteBstWouldAddEndPuncttrue
\mciteSetBstMidEndSepPunct{\mcitedefaultmidpunct}
{\mcitedefaultendpunct}{\mcitedefaultseppunct}\relax
\EndOfBibitem
\bibitem[{\citenamefont{Stavreva et~al.}(2012)\citenamefont{Stavreva, Olness,
  Schienbein, Jezo, Kusina et~al.}}]{Stavreva:2012bs}
\bibinfo{author}{\bibfnamefont{T.}~\bibnamefont{Stavreva}},
  \bibinfo{author}{\bibfnamefont{F.}~\bibnamefont{Olness}},
  \bibinfo{author}{\bibfnamefont{I.}~\bibnamefont{Schienbein}},
  \bibinfo{author}{\bibfnamefont{T.}~\bibnamefont{Jezo}},
  \bibinfo{author}{\bibfnamefont{A.}~\bibnamefont{Kusina}},
  \bibnamefont{et~al.} (\bibinfo{year}{2012}), \eprint{1203.0282}\relax
\mciteBstWouldAddEndPuncttrue
\mciteSetBstMidEndSepPunct{\mcitedefaultmidpunct}
{\mcitedefaultendpunct}{\mcitedefaultseppunct}\relax
\EndOfBibitem
\bibitem[{\citenamefont{Jadach}(2003)}]{Jadach:2002kn}
\bibinfo{author}{\bibfnamefont{S.}~\bibnamefont{Jadach}},
  \bibinfo{journal}{Comput. Phys. Commun.} \textbf{\bibinfo{volume}{152}},
  \bibinfo{pages}{55} (\bibinfo{year}{2003}), \eprint{physics/0203033}\relax
\mciteBstWouldAddEndPuncttrue
\mciteSetBstMidEndSepPunct{\mcitedefaultmidpunct}
{\mcitedefaultendpunct}{\mcitedefaultseppunct}\relax
\EndOfBibitem
\bibitem[{\citenamefont{Kretzer and Schienbein}(1998)}]{Kretzer:1998ju}
\bibinfo{author}{\bibfnamefont{S.}~\bibnamefont{Kretzer}} \bibnamefont{and}
  \bibinfo{author}{\bibfnamefont{I.}~\bibnamefont{Schienbein}},
  \bibinfo{journal}{Phys. Rev.} \textbf{\bibinfo{volume}{D58}},
  \bibinfo{pages}{094035} (\bibinfo{year}{1998}), \eprint{hep-ph/9805233}\relax
\mciteBstWouldAddEndPuncttrue
\mciteSetBstMidEndSepPunct{\mcitedefaultmidpunct}
{\mcitedefaultendpunct}{\mcitedefaultseppunct}\relax
\EndOfBibitem
\end{mcitethebibliography}

\end{document}